\documentclass{article}
\usepackage{spconf,amsmath,graphicx,tabularx}

\title{Improving Performance And Inference On Audio Classification Tasks Using Capsule Networks}
\name{Royal Jain}
\address{AI For All}
\begin{document}
%\ninept
%
\maketitle
\begin{abstract}
Classification of audio samples is an important part of many auditory systems. Deep learning models based on the Convolutional and the Recurrent layers are state-of-the-art in many such tasks. In this paper, we approach audio classification tasks using capsule networks trained by recently proposed dynamic routing-by-agreement mechanism. We propose an architecture for capsule networks fit for audio classification tasks and study the impact of various parameters on classification accuracy. Further, we suggest modifications for regularization and multi-label classification. We also develop insights into the data using capsule outputs and show the utility of the learned network for transfer learning. We perform experiments on 7 datasets of different domains and sizes and show significant improvements in performance compared to strong baseline models. To the best of our knowledge, this is the first detailed study about the application of capsule networks in the audio domain.
\end{abstract}
\section{Introduction}
\label{sec:intro}
Audio Classification tasks are important as part of the complex auditory systems or as the end goal itself. Noise and music event detection is an important pre-processing step for many downstream systems. The performance of speech recognition systems can be improved by supplementing it with a hierarchical gender identification classifier \cite{Bhukya2018EffectOG}. Detecting depression \cite{10.1007/978-3-319-76908-0_37} and emotion \cite{Sarma2018} from audio recordings has been getting plenty of attention in recent years due to applications in medical and conversational AI systems.  

Deep Learning models have shown significant improvement over traditional machine learning algorithms in audio classification systems. Convolutional and Recurrent Neural Networks are used for many audio classification tasks \cite{DBLP:journals/corr/CakirPHHV17} \cite{high-level-feature-representation-using-recurrent-neural-network-for-speech-emotion-recognition}. Addition of Attention Layer to recurrent networks  to focus on the most salient parts of an input via weighting has been very successful in numerous applications, including machine translation \cite{DBLP:journals/corr/BahdanauCB14} and image captioning \cite{DBLP:journals/corr/XuBKCCSZB15} and sound event detection \cite{DBLP:journals/corr/XuKHWP17a}
\cite{DBLP:journals/corr/abs-1710-00343}.

The notion of a capsule was first introduced in \cite{10.1007/978-3-642-21735-7_6} and
very recently revisited in \cite{DBLP:journals/corr/abs-1710-09829} with the addition of a routing-by-agreement mechanism. The authors of \cite{DBLP:journals/corr/abs-1710-09829} found that routing-by-agreement with capsules performed better than the state-of-the-art for digit recognition from images.

In this paper, we propose a capsule network architecture (CapsNet) fit for audio classification tasks and compare it with other deep learning models. We experimentally analyze the impact of various parameters on performance  and introduce changes to loss function and network architecture for adding regularization and multi-label classification. We also present insights into the data using the output of the capsules and show utility of the trained network for transfer learning.

\begin{figure*}[h]
\centering
\includegraphics[width=16.0cm, height=4.0cm]{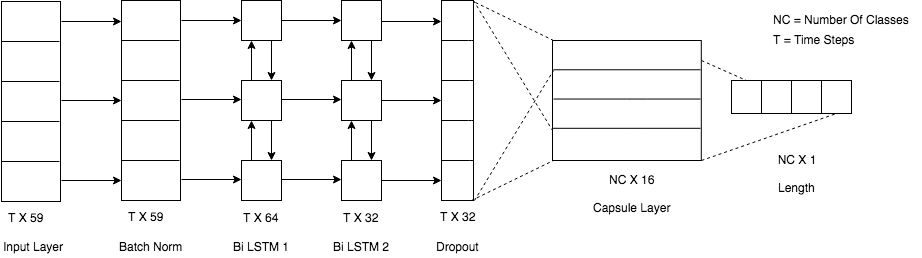}
%  \vspace{2.0cm}

\caption{Capsule Network Architecture}
\end{figure*}

\section{Background Work}
\label{sec:format}

Deep learning approaches are becoming popular for modelling emotion-specific information from speech signals \cite{speech-emotion-recognition-using-deep-neural-network-and-extreme-learning-machine}. Adding attention mechanism to Bidirectional LSTM models improves the performance significantly \cite{Sarma2018}. \cite{QAWAQNEH20175} obtained good results on gender and age identification tasks using deep learning approaches. \cite{DBLP:journals/corr/SalamonB16} showed
superior performance of deep convolutional networks on sound event detection task.

While these approaches have worked well, they have some major drawbacks. CNNs have ability to deal with translation, but for the other dimensions of an affine transformation we have to choose between replicating feature detectors on a grid that grows exponentially with the number of dimensions, or increasing the size of the labelled training set \cite{DBLP:journals/corr/abs-1710-09829}. RNNs/LSTMs, apart from being much slower than CNNs, are highly dependent on the temporal position of entities. Hence, low variability on the temporal scale in training data can deteriorate the generalizability of the models. We intend to overcome these drawbacks using capsule network architecture and dynamic routing-by-agreement mechanism. 

A capsule is a group of neurons whose output vector represents the instantiation
parameters of a specific type of entity such as an object. The length of the output vector represents the probability that the entity exists and its direction represents the orientation of the entity. Active capsules predict the instantiation parameters of higher-level capsules, via transformation matrices. Routing-by-agreement mechanism enables a lower-level capsule to send its output to higher level capsules whose activity vectors have a big scalar product with the prediction advancing from the lower-level capsule. The motivation for capsule routing is that it implicitly learns global coherence by enforcing part-whole relationships to be learned. Capsule Networks and routing-by-agreement mechanism is described in detail in \cite{DBLP:journals/corr/abs-1710-09829}.

In recent times, capsule networks have been applied to a variety of image classification problems like brain tumour detection \cite{Afshar2018BrainTT}, traffic sign detection \cite{Kumar2018NovelDL}. \cite{Iqbal2018CapsuleRF} have used capsule networks for Sound Event Detection task.

\section{Datasets}
\label{sec:format}

For our experiments, we use datasets across different domains and sizes to capture the applicability of proposed system for general use. We now describe the datasets in detail.

\begin{itemize}

\item {\bf IEMOCAP} : The IEMOCAP dataset proposed in \cite{Busso2008IEMOCAPIE} is a multi-modal emotion recognition dataset containing improvised and scripted scenarios designed to elicit specific types of emotions. We use audio data of 4 emotion categories as per existing research \cite{Sarma2018}.

\item {\bf RAVDESS} : The RAVDESS dataset proposed in \cite{10.1371/journal.pone.0196391} is a multi-modal database of emotional speech and song. We use speech data representing calm, happy, sad, angry, fearful, and neutral expressions.

\item {\bf VCTK} : The VCTK corpus \cite{Veaux2017} contains utterances by 109 speakers. We use this dataset for gender identification from sound. The dataset has 47 male speakers and 61 female speakers and 1 unspecified speaker.

\item {\bf Urban8K} : Urban8K corpus \cite{Salamon:UrbanSound:ACMMM:14} contains 8732 labelled sound excerpts of urban sounds from 10 classes like air\_conditioner, car\_horn, children\_playing etc

\item {\bf FSDD} : FSDD is a small dataset \cite{zohar_jackson_2018_1342401} containing spoken utterances of 10 digits by three speakers out of which one speaker is used only for testing. We calculate the classification accuracy for 10 classes for the speaker not used in training.

\item {\bf Multi-VCTK} : We create an augmented dataset (Multi-VCTK) from VCTK corpus which contains the concatenation of male(M) and female(F) speakers thereby creating four categories of M-M, M-F, F-M and F-F.
 
\item {\bf Multi-MUSAN} : The MUSAN corpus \cite{DBLP:journals/corr/SnyderCP15} contains utterances of music, speech, and noise. We create "Multi-Musan" corpus by augmenting some of the speech files with noise files with SNR 5. The resultant dataset contains three categories of speech, noise and speech-noise.

\end{itemize}

\begin{table*}[t]
\centering
\noindent\begin{tabularx}{\textwidth}{|X|l|l|l|l||l|l|l||l|l|}
 \hline
  Datasets & CNN & LSTM & ATT & CAPS & 1 & 3 & 5 & No Regularization & Regularization\\ 
  \hline
 VCTK & 0.797 & 0.895 & 0.894 & {\bf 0.905} & {\bf 0.925} & 0.902 & 0.900 & {\bf 0.925} & 0.902 \\ 
 IEMOCAP & 0.512 & 0.530 & 0.566 & {\bf 0.583} & 0.586 & 0.586 & {\bf 0.587} & 0.579 & {\bf 0.582}\\ 
 Urban8K & 0.518 & 0.585 & 0.595 & {\bf 0.662} & {\bf 0.695} & 0.669 & 0.684 & 0.636 & {\bf 0.650}\\
 FSDD & {\bf 0.728} & 0.673 & 0.689 & 0.692 & {\bf 0.760} & 0.695 & 0.651 & 0.648 & {\bf 0.730}\\
 \hline
\end{tabularx}

\caption{Model Accuracy}
\label{tab:accuracies}
\end{table*}

\section{Model}
\label{sec:format}

We use series of MFCC features as input to the model.
For MFCC features, the scale of different features varies considerably over an input, this leads to longer time for convergence and numerical instability. We add a batch normalization layer to counter the large variability in scale of inputs. Experimentally we've found, for audio data, adding LSTM layer before the capsule layer is much more beneficial than convolutional layers. Hence, we add two Bidirectional LSTM layers. The outputs of each cell of LSTM is passed to the next layer. To improve generalization we add a Dropout Layer. The output of this layer is treated as a series of vectors which is passed to capsule layer. We use capsules of dimension 16. Then we add  Length layer which calculates the length of the activity vector of capsule layer which denotes the probability of the class being present. This model is depicted in figure 1. The architecture of the model is inspired by \cite{DBLP:journals/corr/abs-1710-09829}, however, unlike \cite{DBLP:journals/corr/abs-1710-09829} we don't add decoder network except in experiments where we test the utility of regularization. This is done to ensure a fair comparison with other models. We use margin-loss as described in \cite{DBLP:journals/corr/abs-1710-09829} as the loss function.

\begin{equation} \label{eq:1}
L_{k} = T_{k}max(0, m^{+} - ||v_{k}||)^{2} +
\lambda(1 - T_{k})max(0, ||v_{k}|| - m^{-})^{2}
\end{equation}

where $T_{k}$ = 1 if class k is present
and $m^{+}$ = 0.9 and $m^{-}$ = 0.1. The $\lambda$ down-weighting of the loss for absent classes stops the initial learning from shrinking the lengths of the activity
vectors of all the capsules. We use $\lambda$ = 0.5. 

Apart from Capsule Network, we use Deep Convolutional Networks (CNN), Bidirectional LSTM networks (LSTM) and Bidirectional LSTMs with attention mechanism (ATT) as baselines in our experiments.

%\subsection{CapsNet}
%\label{ssec:subhead}

\section{Experiments}
\label{sec:format}

We experiment with model architectures presented in the previous section. The effect of the dimensionality of capsule output, number of routings in the routing-by-agreement mechanism and regularization on the performance of capsule networks are analyzed. Thereafter, we evaluate the models under multi-label setting. For our experiments, we use the standard MFCC features of 40ms frame length and 10ms hop length. For each frame, we calculate 19 leading coefficients along with delta and  delta-delta derivatives and energies. Although adding other features such as zero-crossing rate (ZCR), root mean square (RMS) energy, pitch etc. can give better results than using MFCCs alone. We refrain from including them as they are task specific and the goal is to compare different models and parameter values in general setting rather than getting the best results on some particular task. All the datasets are approximately balanced, hence we choose accuracy as our performance metric.

\subsection{Model Comparison}
\label{ssec:subhead}

As shown in Table \ref{tab:accuracies}, we observe significant improvement in performance with capsule networks when compared with other deep learning models, except for FSDD dataset. The capsule network consistently out-performs Attention LSTM network which is considered state-of-the-art for many audio classification tasks. The improvement is especially prominent in the Urban8K dataset where accuracy improved from 59.5\% to 66.2\%. This experiment shows the advantage of using the capsule network with dynamic routing-by-agreement compared to other models.

\subsection{Effect of Routings}
\label{ssec:subhead}

We experiment with three values of routings 1,3 and 5. Like \cite{DBLP:journals/corr/abs-1710-09829}, we found that higher values of routings tend to overfit the dataset. In Table \ref{tab:accuracies}, we can see that this is especially noticeable for a small dataset like FSDD.

\begin{figure}[htb]

\includegraphics[width=8.0cm]{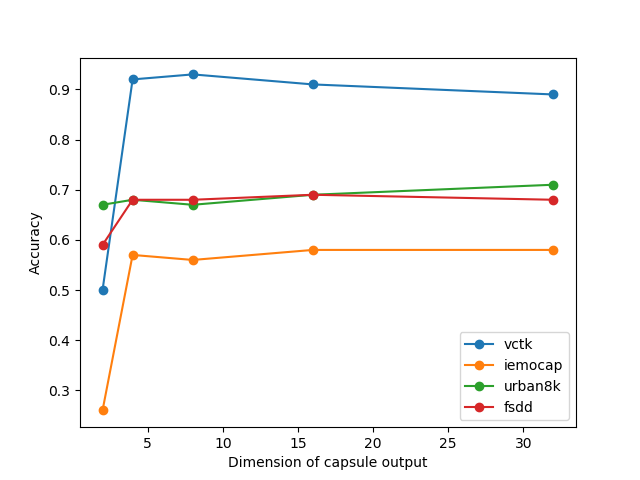}
%  \vspace{2.0cm}

\caption{Dimension of Capsule vs Accuracy}
\label{fig:dimnesion}
\end{figure}

\subsection{Effect of Dimensionality of Capsules}
\label{ssec:subhead}

We vary the dimension of capsules from 2 to 32 and measure the accuracy on the test set. From figure \ref{fig:dimnesion}, we observe a trend in accuracy that it first increases with dimension and then decreases. One plausible explanation of this behaviour is that, at lower dimensions, CapsNet doesn't have the capacity to learn and it underfits the data, while at higher dimension it becomes easier to overfit the training data. Thus, the model capacity acts as a regularizer. One more supporting evidence is that datasets with lesser data (VCTK) are more affected by increasing the dimension.

\subsection{Regularization}
\label{ssec:subhead}

Inspired by \cite{DBLP:journals/corr/abs-1710-09829}, we increase regularization by adding a decoder network of fully connected layers after the capsule layer to reconstruct the MFCCs. To reconstruct the inputs, the capsules have to encode the instantiation
parameters of the input signal. As the scale of the data varies across features, we first scale the data using min-max scaling to prevent few features dominating the loss. Total loss for the model is the weighted sum of margin loss and Mean Absolute Error for reconstruction. As seen in Table \ref{tab:accuracies}, adding regularization is generally beneficial especially when training data is not very large like FSDD.

\begin{table}
\centering
\begin{tabular}{ |c|c|c| } 
 \hline
  & ATT & CAPS \\ 
  \hline
 Multi-VCTK & 0.877 & {\bf 0.90}  \\ 
 Multi-MUSAN & 0.721 & {\bf 0.737} \\ 
 \hline
\end{tabular}
\caption{Multi-label Accuracy}
\label{tab:multilabel}
\end{table}

\subsection{Multi-label Setting}
\label{ssec:subhead}

In this setting, we compare Bidirectional LSTM with Attention model, trained using binary cross-entropy loss, with Capsule Networks. We use  Multi-VCTK and Multi-Musan datasets described in section 3. Weighted accuracy is used for comparing the models. We change the weighing down parameter, $\lambda$, from equation \ref{eq:1} to 1 to increase the effect of false positives. Results in Table \ref{tab:multilabel} suggest that routing-by-agreement allows each capsule to pick inputs which are consistent with its current output by increasing their weights and decreasing the weight of others. This reduces interference of other classes, leading to better results.

\section{Inference \& Transfer Learning}
\label{sec:format}

\begin{figure}[htb]

\includegraphics[width=7.0cm]{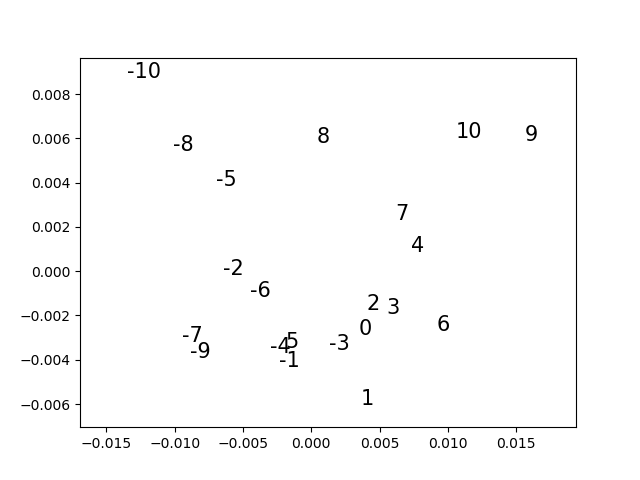}
%  \vspace{2.0cm}

\caption{Scatter Plot of Amplitude Variations}
\label{fig:amp}
\end{figure}

\begin{figure}[htb]

\includegraphics[width=7.0cm]{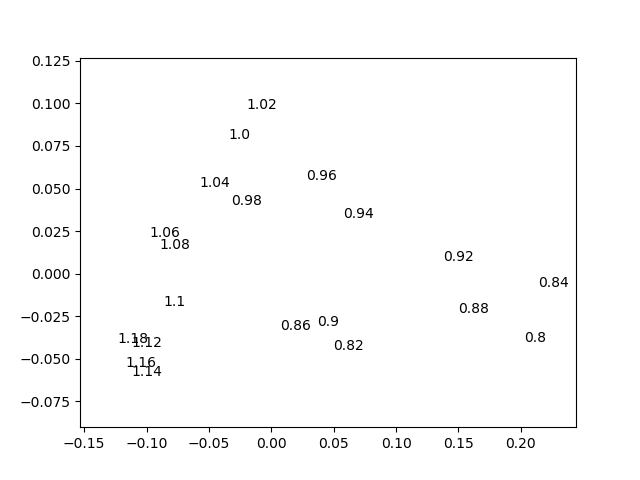}
%  \vspace{2.0cm}

\caption{Scatter Plot of Speed Variations}
\label{fig:speed}
\end{figure}

\subsection{Amplitude \& Speed Variation}
\label{ssec:subhead}
In \cite{DBLP:journals/corr/abs-1710-09829}, the authors infer the effect of coefficients of the capsule's activity vector by perturbing the vector by a small amount and observing the reconstructed output. MFCCs lose a lot of information about voicing/pitch and thus it is not possible to generate good quality audio back from MFCCs. Hence, unlike images, observing the effect by perception is not the ideal solution. Instead, we alter the input signal slightly and observe the changes in capsule output.

We vary the amplitude by adding and reducing constant signals. The CapsNet model is trained on this augmented data. We then pass the same audio with different values of the added signal and observe the capsule activity vectors by applying PCA on the output of the relevant capsule. This is plotted in figure \ref{fig:amp}. The label denotes the fixed signal used for augmentation. We perform a similar experiment with augmented data generated by reducing and increasing speed. The result is in figure \ref{fig:speed}, where the label signifies the factor by which the signal was sped up. In both figures, we observe that the signals with increased amplitude/speed are clearly separated from those with decreased amplitude/speed. Moreover, two components are able to explain majority of variance in data. This implies that output vector stores information about the orientation(pose) of the input signal and it only requires few dimension in PCA space to store the majority of information about a particular aspect of orientation.

\subsection{Transfer Learning}
\label{ssec:subhead}

Transfer learning applies knowledge gained on the source domain to a related target domain. Capsule Network offers a direct way of applying transfer learning as we can use the activity vectors of capsules as supplementary features. To verify this, we use the model trained on Iemocap dataset and use the output of the capsule layer as additional input along with MFCCs on Ravdess dataset, where the emotion classes and settings are different but related. The six class classification accuracy jumped from 41\% to 50\% with transfer learning signifying that activity vectors of capsule contain information about presence and orientation of corresponding entity which can be translated to a related domain.

\section{Conclusion}
\label{sec:format}

In this paper, we have proposed a novel capsule network architecture and regularization technique fit for the audio domain. We showed that our approach works better than other deep learning approaches under similar conditions. Regularization, along with varying the number of routings and dimension of capsule, can significantly improve the results. We also demonstrated that outputs of capsules can be used for getting insights into the data and for transfer learning on a related dataset.

\newpage

% References should be produced using the bibtex program from suitable
% BiBTeX files (here: strings, refs, manuals). The IEEEbib.bst bibliography
% style file from IEEE produces unsorted bibliography list.
% -------------------------------------------------------------------------
\bibliographystyle{IEEEbib}
\bibliography{strings,refs}

\end{document}